\documentclass[%
 reprint,
 amsmath,amssymb,
 aps,
]{revtex4-2}

\usepackage{graphicx}
\usepackage{dcolumn}
\usepackage{bm}
\usepackage{hyperref}
\usepackage[usenames,dvipsnames]{color}
\usepackage{orcidlink}

\begin{document}

\preprint{APS/123-QED}

\title{Uncovering the bias in the evidence for dynamical dark energy through minimal and generalized modeling approaches}

\newcommand{\orcid}[1]{\orcidlink{#1}}

\author{Ziad Sakr \orcid{0000-0002-4823-3757}}
 \email{ziad.sakr@net.usj.edu.lb}
\affiliation{%
 Faculty of Sciences, Universit\'e St Joseph, Beirut, Lebanon
}%
\affiliation{IRAP, Université de Toulouse, CNRS, CNES, UPS, Toulouse, France}


\date{\today}

\begin{abstract}
In this letter we argue that the commonly adopted CPL parameterisation for the dark energy equation of state (EoS), is biased towards preferring such model over the constant  $w$ while the latter bounds are still compatible with their $\Lambda$CDM values. For that we compare constraints on the EoS parameters $w_0$ and early type $w_a$ (CPL) against those with a late time parameterisation on $w_a$ (hereafter LZ) and the constant $w$ model, using CMB, Supernovae and BAO from DESI datasets. We found, the same as was the case with CPL model, preference for dynamical dark energy within the LZ model, but for values almost symmetrically distributed with respect to their $\Lambda$CDM limits. This is due to the fact that the presence of $w_0$ allows to recast each parametrisation into making it compensate the preference for $w\sim -1$ in the opposite direction. To further test our hypothesis, we fixed  $w_0$ to -1 and followed a minimal investigation approach by considering models that deviates only by one free parameter following the same previous parameterisations, making them now genuinely different from each others. On the other hand, we extend our models to more general ones by, either incorporating parameterisations that group both late and early $w_a$ effects, which we name GEoS model, or by allowing the presence of two dark energy components, fluid alike $\Omega_{\rm DE}$ and constant alike $\Omega_{\rm CC}$. 
We also complement conclusions drawn from the significance levels from the resulting confidence contours with that of the Bayes factor calculation. 
We found that all the variants, except the original CPL or LZ are still compatible with  $\Lambda$CDM, with likelihoods peaking close to $w_0 = -1$, $w_a =  0$, or 0.68 for $\Omega_{\rm CC}$, with the constant $w$ and the late time $w_a$ (minimal LZ) being the ones with the tightest constraints. And although we found that the evidence from CPL or LZ are stronger than those for the more minimal cases, however, the preference increases further for the more generalized parameterisations, instead of penalizing the addition of the extra degrees of freedom, while still staying compatible with $\Lambda$CDM in terms of the significance levels. We conclude that considering CPL model is not sufficient on its own, to test deviations from the simple cosmological constant model, and that it is necessary to conduct further minimal or more general approaches, when confronting models to observations, so that to localize biases, if they exist, or to better understand the outcomes from model testing and inference methods.
\end{abstract}

\maketitle


\section{Introduction and motivations}

Recently, there have been claims about a preference for a dynamical dark energy over a cold dark matter model with a cosmological constant ($\Lambda$CDM) found when using a combination of datasets from Cosmic Microwave Background from the Planck 2018 Mission (hereafter Plk18) \citep{Planck:2018vyg}, and Supernovae (SN) samples from e.g. Pantheon+ \citep{Brout:2022vxf}, and the latest releases of baryonic acoustic oscillations (BAO) signature in galaxy clustering from the Dark Energy Spectroscopic Instrument (DESI) year one Stage IV DESI mission \citep{DESI:2024mwx} (hereafter DESI24). Despite the fact that the Bayes factor, one of the quantitative indices used to asses a model after being confronted to data, remain inconclusive as noted by DESI24 about which model is favored, especially that it was also noted by the DESI collaboration or \cite{Cortes:2024lgw,Patel:2024odo} that the results might be prior dependent, and despite concerns about the validity of combining constraints from discrepant contours coming separately from CMB, SN and BOA \cite{Wang:2024rjd}, and despite the fact that some of the observations in the DESI24 sample are yet to be confirmed as not outliers to the other collected points \citep{Liu:2024gfy} \citep{Sapone:2024ltl}, an issue that has the effect to mitigate the evidence for dynamical dark energy, a result found as well when using instead the BAO local observations from SDSS survey \citep{DESI:2024uvr}, however, already before DESI24 results we could find some preference or at least the dynamical dark energy was still largely allowed by the CMB power spectrum \citep{Planck:2018vyg}, alone or after a combination with constraints from Supernovae datasets. Moreover, despite the above concerns, many subsequent studies to DESI24 have more or less corroborated the dynamical dark energy preference results with several robustness and 'stress' tests \citep{Pang:2024qyh,Luongo:2024zhc,Zheng:2024qzi,Dinda:2024ktd,Giare:2024gpk,Pourojaghi:2024tmw,Chan-GyungPark:2024mlx,Ghosh:2024kyd,Ramadan:2024kmn,Notari:2024rti,Orchard:2024bve,Roy:2024kni,Giare:2024ocw,Chudaykin:2024gol,Hernandez-Almada:2024ost,Tada:2024znt,DESI:2024aqx,Wolf:2024stt,Wolf:2024eph}, though some \citep{Colgain:2024mtg,Gao:2024ily,Notari:2024zmi,Chan-GyungPark:2024brx,Dhawan:2024gqy,Efstathiou:2024xcq,Colgain:2024ksa,Carloni:2024zpl,Lewis:2024cqj,Colgain:2024xqj} have raised additional new concerns.

However, the evidence for dynamical dark energy was mainly obtained using the Chevallier-Polarski-Linder (CPL) parameterisation \citep{Chevallier:2000qy,Linder:2002et} for the equation of state model $w = w_0+(1-a)\,w_a$ which was originally argued that it does not diverge at high redshift (like the one proposed by e.g. \cite{Astier:2000as}) since it has a finite value at high redshift, though other parameterisations exist that could also not suffer from that issue \cite{Efstathiou:1999tm,Jassal:2005qc,Barboza:2008rh,Pan:2019brc}. Indeed, \cite{Giare:2024gpk} explored and studied the impact of such models in the light of latest DESI24 BAO release but also found similar preference for dynamical dark energy. Despite the good features of CPL, we argue however that it suffers, along with other similar behavior from the similar aforementioned models,  from hindering degeneracies at high redhsift that could bias the results, since practically $w \sim w_0+w_a $ when $a\rightarrow0$. This essentially makes CMB less efficient in breaking the degeneracy relying on strong constraints at low redshifts for doing so. And since the constraints for the constant parameterisation for $w$ has yielded a preference for -1, the equivalent value for the $\Lambda$CDM model, one has to wonder whether the $w_a$ exploration of values $\neq 0$ along with $w_0$ deviating from -1 is not induced by the freedom caused by the increase of parameters, such that, at low redshift, where $w_a$ has little effect, $w_0$ is allowed to go to $<-1$, then staying so till high redshifts where a negative value for $w_a$ will compensate a $w_0$ shift of $w$ from its $w = -1$ preference. This could be hinted from the work of \cite{Ye:2024ywg} where we see a symmetric oscillation of a constructed $w_{\rm DE}$ around -1 (see also \cite{DESI:2024kob} for similar behavior but with physics-focused dynamical dark energy models). In order to test this intuition, we propose to repeat the same study here but with another parameterisation, almost with counter behavior, in terms of dynamical evolution and signature in the three above probes, dubbed late redshift (LZ) paramerisation $w = w_0 + a  \, w_a$ that also does not diverge at high redshift.

We will also conduct other stress and robustness tests, that could also allow us to better understand our results, such as a minimal investigation approach by considering models that deviates only by one free parameter within the same previous parameterisations, or we extend our models to more general ones by, either incorporating parameterisations that group both CPL and LZ ones, which we name GEoS model, alone or by further allowing the presence of two dark energy components, fluid alike $\Omega_{\rm DE}$ and constant alike $\Omega_{\rm CC}$.

This paper is organized as follows: we begin in Sect.~\ref{sec:model} by presenting the different parameterisation we shall test in this work, and continue in Sect.~\ref{sec:datamod} by presenting the pipeline and data used in our analysis, to then show and discuss our results in Sect.~\ref{sect:result}, before concluding in Sect.~\ref{sect:conclusions}. 
\section{From minimal to more general extensions}\label{sec:model}

In this section, we review the simple standard cosmological model framework first (just for completeness) and then introduce further extensions beyond the $\Lambda$CDM model, which is defined as the cold dark matter with dark energy accounted as a cosmological constant in the following Einstein's field equations 
\begin{equation}
G_{\mu \nu} + \Lambda g_{\mu \nu} = \kappa T_{\mu \nu}      \label{eq:EFE}
\end{equation}
where $G_{\mu \nu}$ is the Einstein tensor, $g_{\mu \nu}$ is the metric tensor, $T_{\mu \nu}$ is the stress–energy tensor, $\Lambda$ is the cosmological constant and $\kappa=8\pi G$ is the Einstein gravitational constant (we work in speed of light $c =1$ units).
Assuming the homogeneous and isotropic universe described by the Friedmann-Lema\^{i}tre-Robertson-Walker metric (FLRW) metric 
\begin{equation}
\mathrm{d}s^2=-dt^2+a^2(t)\left[\frac{\mathrm{d}r^2}{1-kr^2}+r^2\mathrm{d}\theta^2+r^2sin^2\theta \mathrm{d}\phi^2\right],      \label{eq:FRW}
\end{equation}  
where $a(t)$ and $k$ are the scale factor at cosmic time $t$ and the curvature of spacetime, respectively. Substituting Eq.~\ref{eq:FRW} into the Einstein's field equations \ref{eq:EFE}, we obtain, in the Universe spatial flat assumption, the so-called Friedmann equations as follows
\begin{equation}
H^2(a)=\frac{\dot{a}}{a}=\frac{8\pi G}{3}\sum^i\rho_i,     \label{eq:f1}
\end{equation}   
\begin{equation}
\dot{H}(a)+H(a)=\frac{\ddot{a}}{a}=-\frac{4\pi G}{3}\sum^i(\rho_i+3p_i),     \label{eq:f2}
\end{equation}   
where $H(a)$ is the Hubble parameter, and $\rho_i$ and $p_i$ represent the volume energy density and pressure of different species in the cosmic budget. Eq.~\ref{eq:f1} can be written, with $a(t)=(1+z)^{-1}$, $z$ being the observed redshift of the source and the same symbol $H$ re-used, 
\begin{equation}
\frac{{H^2(z)}}{{H^2(z=0)}}=\frac{8\pi G}{3}\sum^i\,\Omega_i(z),    
\end{equation} 
where $\Omega_i(z)$ is the energy density.
A quantity that will enter the theoretical formulation of many of the observables we use in this work is $E(z)$, the dimensionless Hubble parameter, that depicts the background evolution of a specific cosmological model, obtained after combining Eq.~\ref{eq:f1} and \ref{eq:f2}, for the $\Lambda$CDM scenario, 
\begin{equation}
E_{\mathrm{\Lambda CDM}}(z)=\left[\Omega_{\rm M}(1+z)^3+\Omega_{\rm R}(1+z)^4+{\Omega_\Lambda} \right]^{\frac{1}{2}}, \label{eq:ezlcdm}
\end{equation}
where $M$, $R$ and $\Lambda$ stands (at present time value $z=0$) for matter, radiation and the symbol for dark energy as a constant in Einstein's field equations  \ref{eq:EFE}, with $(\Omega_{\rm DE}=1-\Sigma \, \Omega_i)$;
or for a common and effective way to describe an extension of $\Lambda$CDM, we choose a dark energy equation of state parameter $w=\frac{p_{DE}}{\rho_{DE}}$ constant (with $w =-1$ giving $\Lambda$CDM) called the $w$CDM model , where $\Omega_{\rm DE} = 1- \Omega_{\rm M}$ 
\begin{equation}
E_{\mathrm{wCDM}}(z)=\left[\Omega_{\rm M}(1+z)^3+\Omega_{\rm R}(1+z)^4+{{\Omega_{\rm DE}}^{3(1+w)}} \right]^{\frac{1}{2}}. \label{eq:ezwcdm}
\end{equation}
We could also consider further evolution of the dark energy equation of state, through the CPL phenomenological parametrisation \cite{Chevallier:2000qy,Linder:2002et}. 
\begin{equation} 
 w(z)=w_0 +\frac{z}{1+z} \,w_a
 \label{eq:CPL}
\end{equation}

Its $E(z)$ reads as (neglecting radiation here for display purposes but accounted in our codes later):
\begin{equation}
E_{\mathrm{CPL}}(z)=\left[\Omega_{\rm M}(1+z)^3+\Omega_{\rm DE}(1+z)^{3(1+w_0+w_a)}\mathrm{e}^{\frac{-3w_az}{1+z}}\right]^{\frac{1}{2}}. \label{eq:ezcpl}
\end{equation}
%
Furthermore, to study the dark energy evolution, we consider a new parameterisation with late time effect (hereafter LZ), which we note that it could as well be used later after fixing $w_0$, and the same for the previous CPL.
\begin{equation} 
 w(z)=w_0 +\frac{1}{1+z} \,w_a
 \label{eq:LZS}
\end{equation}

 Its $E(z)$ reads as: 
\begin{equation}
E_{\mathrm{LZ}}(z)=\left[\Omega_{\rm M}(1+z)^3+\Omega_{\rm DE}(1+z)^{3(1+w_0)}\mathrm{e}^{\frac{3w_a z}{1+z}}\right]^{\frac{1}{2}}. \label{eq:ezcpl}
\end{equation}

To go further we shall later consider a more generalized dark energy equation of state (GEoS) evolution grouping the early and late models, for which we must now differentiate the symbol used for $w_a$
\begin{equation} 
 w(z)=w_0 +\frac{z}{1+z} \,w_{a,\rm E}+\frac{1}{1+z} \,w_{a,\rm L},
 \label{eq:GEoS}
\end{equation}
and we go even further with a variant in which we keep $\Omega_\Lambda$ along with $\Omega_{\rm DE}$ which evolution is defined through $w(z)$, but where $\Omega_\Lambda$ is a free parameter now and $\Omega_{\rm DE} = 1 - \sum^i  \Omega_i -\Omega_\Lambda$ obtained from the new closure equation so that its $E(z)$ reads as: 
\begin{equation}
E_{{f(w)+ \rm CC}}(z)=\left[\Omega_{\rm M}(1+z)^3+\Omega_{\rm CC}+\Omega_{\rm DE}(1+z)^{f[w(z)]} \right]^{\frac{1}{2}}. \label{eq:ez+CC}
\end{equation}
Where $f[w(z)]$ would take different values following one of the three parameterisations we considered above. Note that we name in this case and in this work $\Omega_\Lambda$ as $\Omega_{\rm CC}$ although they are the same just to highlight that it will be constrained within a frame of an extension of the original $\Lambda$CDM model. We also note that $\Omega_{\rm CC}$ in this case could take negative values providing $\Omega_{\rm (effective)}= \Omega_{\rm CC} + \Omega_{\rm DE}$ stays positive as considered in different works \cite{Mukhopadhyay:2007ed,Menci:2024rbq,Wang:2024hwd} driven by theoretical motivations from string theory as argued in \cite{Akarsu:2019hmw}, but we choose in this work to limit its values between 0 and 1 because we want to keep the minimal approach that takes  $\Lambda$CDM or $(w=-1)$CDM as our basic models to be extended.
\section{Analysis and datasets}\label{sec:datamod}
For our analysis, we consider CMB angular power spectrum observations from the Plk18 temperature (TT) likelihood, polarization (EE) and their cross-correlation (TE) data using the high-$\ell$ \texttt{plik-lite}, a compressed and faster high-$\ell$ likelihood of \texttt{plik}, which already includes marginalization over foregrounds and residual systematics and is enough accurate for our purpose, and the low-$\ell$ TT \texttt{Commander} and \texttt{SimAll} EE likelihoods \citep{Planck:2019nip}. We complement CMB probe with the Planck lensing likelihood \cite{Planck:2018lbu}. We also combine with the baryonic acoustic oscillations BAO in galaxy power spectrum at different redshifts from DESI24 measurements as specified in \cite{DESI:2024mwx,DESI:2024lzq,DESI:2024uvr}, which includes the BGS sample in the redshift range $0.1 < z < 0.4$, LRG samples in $0.4 < z < 0.6$ and $0.6 < z < 0.8$, combined LRG and ELG sample in $0.8 < z < 1.1$, ELG sample in $1.1 < z < 1.6$, quasar sample in $0.8 < z < 2.1$ and the Lyman-$\alpha$ Forest Sample in $1.77 < z < 4.16$.  Finally, we additionally use with the luminosity distances from type Ia Supernovae probe as compiled in the Pantheon+ SN sample \citep{Scolnic:2021amr} and the corresponding public likelihood from \cite{Brout:2022vxf} . We modified the cosmological and Boltzmann solver \texttt{CLASS} to implement the different models described in Sect.~\ref{sec:model} and use \texttt{MontePython}, the Monte Carlo code \citep{Brinckmann:2018cvx} to estimate our parameters, in which we adapted when necessary the used likelihoods.

\begin{figure}[t]
	\centering
	\includegraphics[width=1.0\columnwidth]{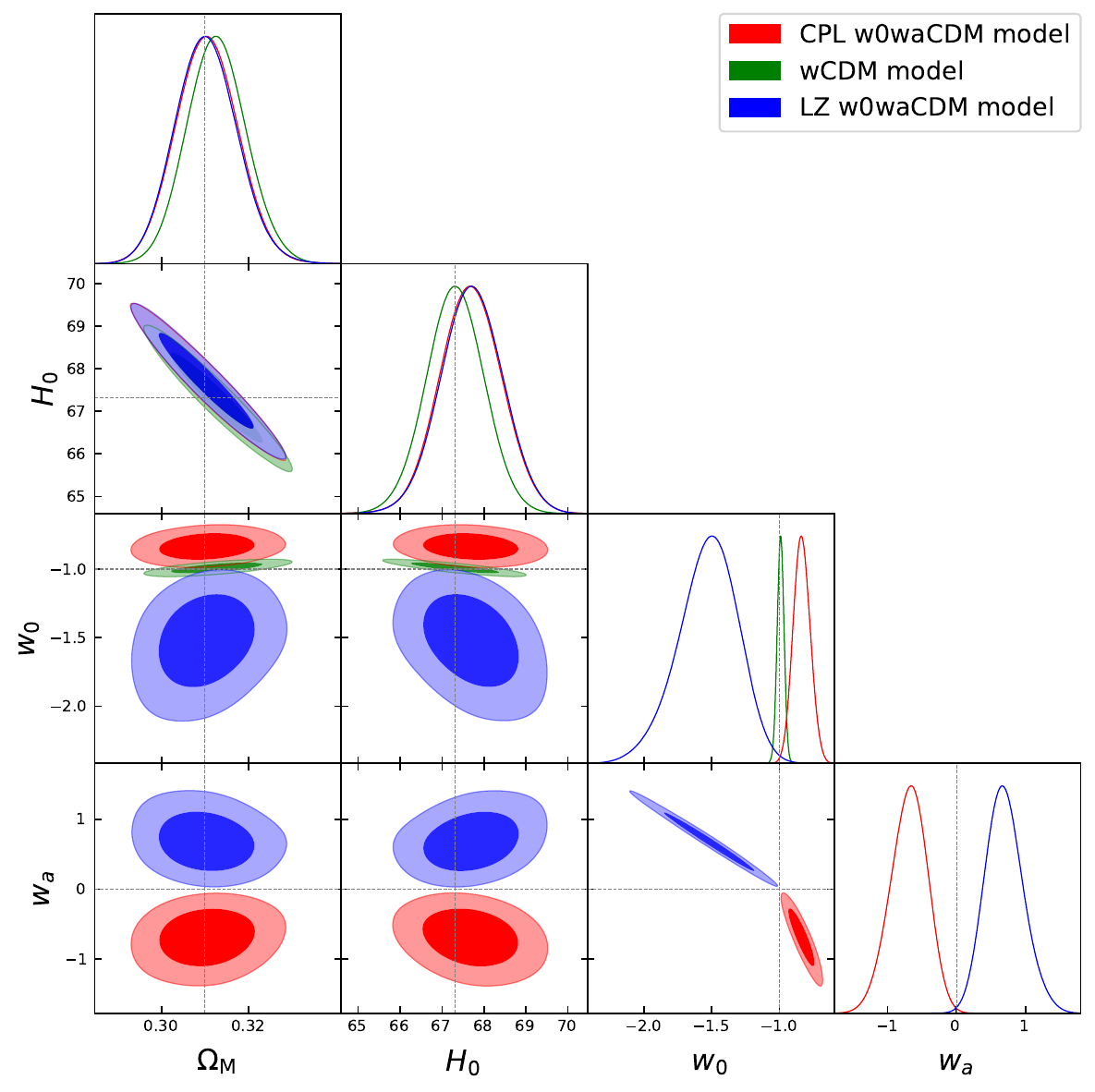}    
	\caption{ The 1D and 2D 68\% and 95\% confidence contours marginalised likelihood for the cosmological parameters $\Omega_{\rm M}$, $H_0$ and the dark energy equation of state parameters $w_0$ and $w_a$ following the two models, early type CPL and late type LZ for dark energy evolution defined in Section~\ref{sec:model}, all inferred from a combinations of CMB correlations from Plk18, Supernovae sample from Pantheon+, and BAO measurements from DESI24. }
	\label{fig:w0wanogeneralisedW}
\end{figure}
%
\section{Results and discussion}\label{sect:result}

We start by showing in Fig.~\ref{fig:w0wanogeneralisedW}, confidence contours inferred from our MCMC following the method described in Sect.~\ref{sec:datamod} using a combinations of CMB correlations from Plk18, Luminosity distances from SN Pantheon+ sample, and BAO measurements from DESI24, and that for the cosmological parameters $\Omega_{\rm M}$, $H_0$ and the dark energy equation of state parameters $w_0$ and $w_a$. We consider two models, early type CPL and late type LZ for the dark energy EoS evolution defined in Section~\ref{sec:model} along with the constant $w$ model. 
We first recover the result of \cite{DESI:2024mwx} when using the same CPL parameterisation, with preference for values of $w_0$, $w_a$ far from their $\Lambda$CDM limits. However, we also find as well preference for dynamical dark energy when using the new parameterisation we introduced in this work, but for values shifting to opposite and almost symmetric space of variation, with respect to, either $w_0$ if we take the constraints on $w$ constant as a reference, or to $w_a$ contours with respect to their null values, an observation that could be explained by the fact that each model is biased towards a direction in the space of parameters such that to adjust for the loss of concordance found around $w=-1$. The explanation of such behavior resides in the way that each model is impacting either early probes such as the CMB datasets or late probes such as the BAO or Supernovae sample. As so, in the CPL parameterisation at low redshift, where $w_a$ has little effect, $w_0$ is allowed to go to $<-1$, then staying so till high redshifts where a negative value for $w_a$ will compensate a $w_0$ shift from its $-1$ preference. The opposite occurs for LZ parameterisation where $w_0$ is allowed to go to $>-1$ at high redshift and stays like that, since it is a constant by construction at low redshift where $w_a$ now compensates by going to positive values.

\begin{figure}
	\centering
          \includegraphics[width=0.9\columnwidth]{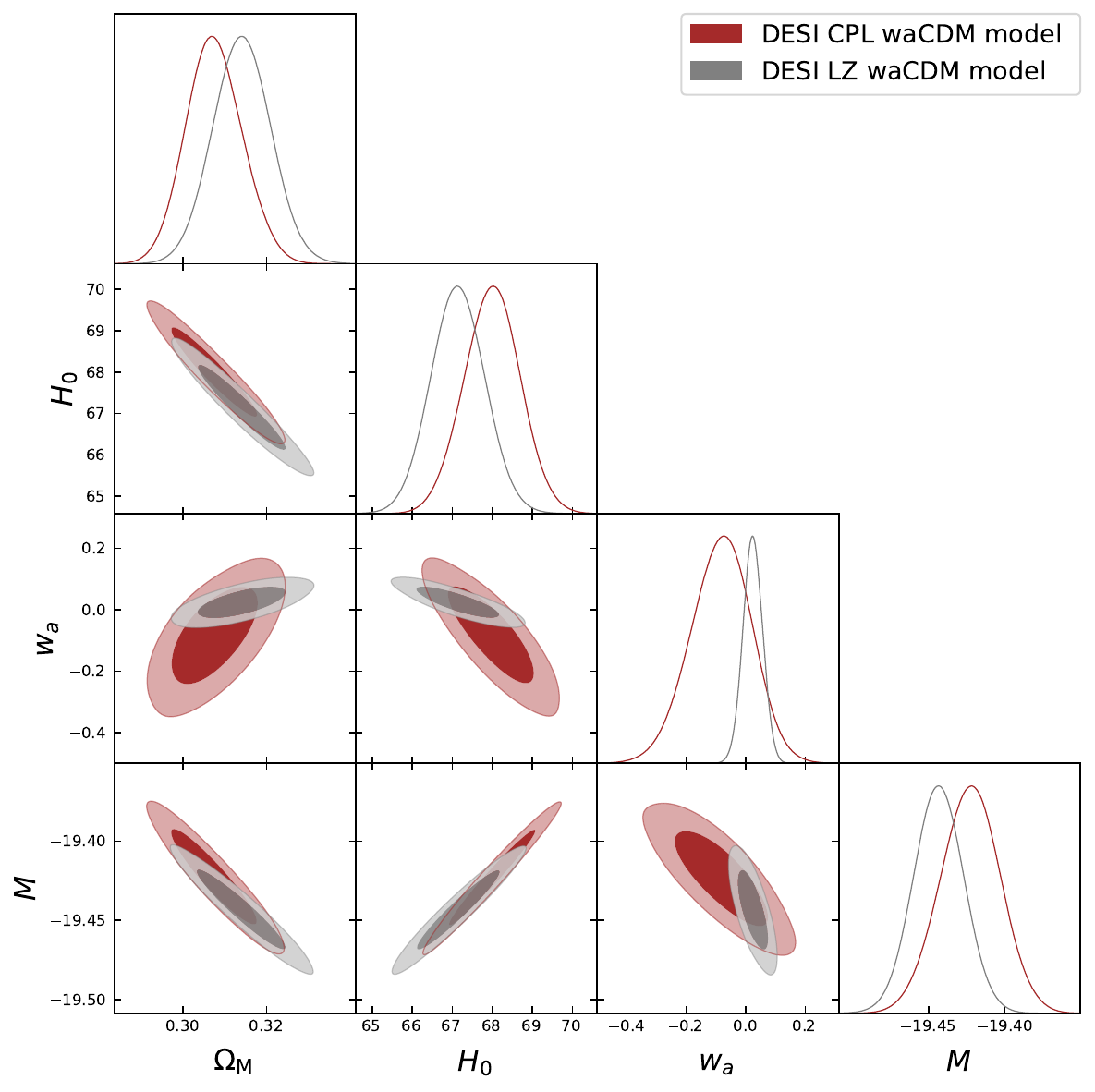}
          
	\caption{The 1D and 2D 68\% and 95\% confidence contours marginalised likelihood for the cosmological parameters $\Omega_{\rm M}$, $H_0$ and the dark energy equation of state parameters $w_0$ and $w_a$ following the two models, minimal early type CPL and minimal late type LZ for dark energy evolution defined in Section~\ref{sec:model}, all inferred from a combinations of CMB correlations from Plk18, Supernovae sample from Pantheon+, and BAO measurements from DESI24. 	
	}
	\label{fig:minimalwanogeneralisedW}
\end{figure}

To further test our hypothesis, we show in Fig.~\ref{fig:minimalwanogeneralisedW} parameter constraints following minimal approaches, where we fix the value of $w_0$ to its $\Lambda$CDM limit and investigate variants where $w$ is parameterised by only one free parameter $w_a$ with late or early type effect, i.e what we can call a minimal LZ or a minimal CPL model, where both are now totally different from each others, and, unlike CPL and LZ, are also different in cc for all the other cosmological, or calibration parameters. We found then, that both $w_a(s)$ are showing no statistically significant shifting and are fully compatible with their null values, with the strongest constraints, almost equal in range to those for $w$ constant case, happening for $w_a$ in the minimal LZ parameterisation. 

\begin{figure}
	\centering
          \includegraphics[width=0.65\columnwidth]{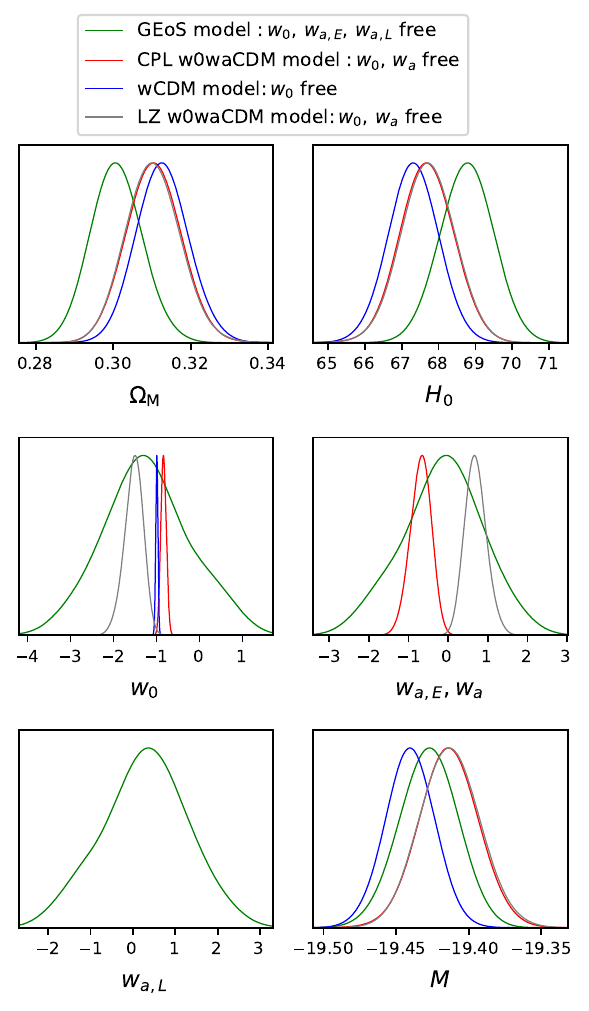}
          
	\caption{The 1D marginalized likelihood for the cosmological parameters $\Omega_{\rm M}$ and $H_0$, the fluid-like dark energy equation of state parameters $w_0$, $w_{a,\rm E}$, and $w_{a,\rm L}$ following the GEoS model defined by Eq.~\ref{eq:GEoS}, along with the same two previous EoS models, the dark energy equation of state parameters $w_0$ and $w_a$ following the two models, early type CPL and late type LZ, all inferred from a combinations of CMB correlations from Plk18, Supernovae sample from Pantheon+, and BAO measurements from DESI24.}
	\label{fig:minimal_generalisedW1DnoMbisprio}
\end{figure}

\begin{figure}
	\centering
          \includegraphics[width=\columnwidth]{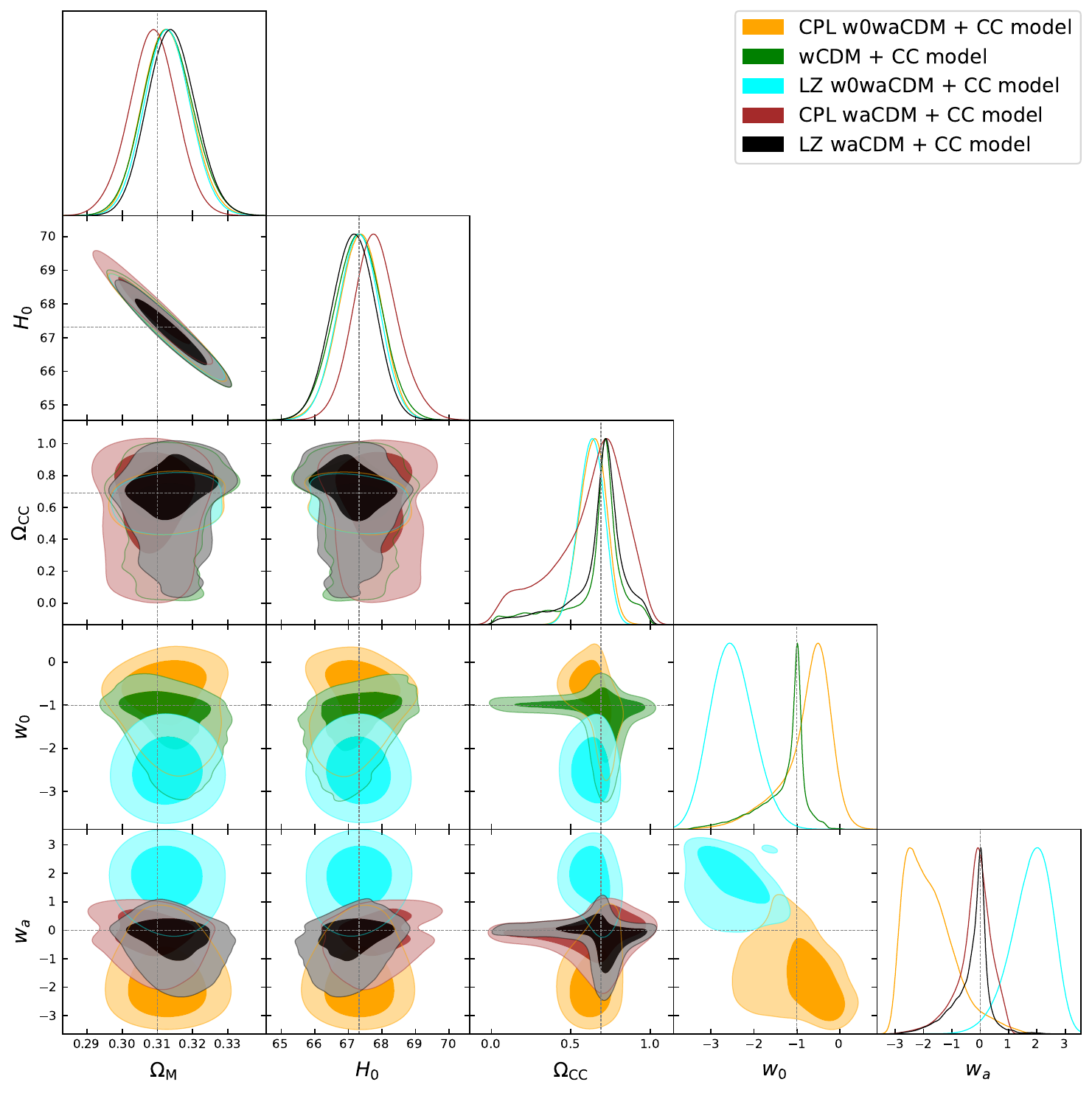}
          
	\caption{The 1D and 2D 68\% and 95\% confidence contours marginalised likelihood for the cosmological parameters $\Omega_{\rm M}$ and $H_0$, the fluid-like dark energy equation of state parameters $w_0$, $w_a$ following the two models, early type CPL and late type LZ , along with the same previous model but with further addition of a cosmological constant alike dark energy component as described in Section~\ref{sec:model}, all inferred from a combinations of CMB correlations from Plk18, Supernovae sample from Pantheon+, and BAO measurements from DESI24. Note that we use the same notation and parameter contour presentations here for either $w_a$ or $w_{a,\rm L}$.}
	\label{fig:minimalw0wanogeneralisedWCCnoMprio}
\end{figure}

To better understand the contribution and role of each model we show in Fig.~\ref{fig:minimal_generalisedW1DnoMbisprio} a comparison with the more general parameterisation (GEoS) as introduced in Sect.~\ref{sec:datamod}. Note that in this parameterisation we introduced a further additional free parameter so that each of the late or early behavior is supposed to be caught by either $w_a$ (early) or $w_{a,\rm L}$ (late). We observe then that the maximum of the likelihood falls around the $\Lambda$CDM limit value for all our parameters though with a widening of the constraints due to the fact, as we shall see later, that the presence of $w_0$ permits to recast the parameterisation into one with only two degrees of freedom. Notice how $\Omega_{\rm M}$ and $H_0$ are shifted and compensated with respect to the late or early parameterisation. On the contrary, notice how the Supernovae calibration parameter $M$ is centred between the two likelihood contours of either the two late or early parameterisation. We see by then that in GEoS there exist a shift in the parameters, however not in the EoS ones, and unlike what we see for EoS parameters for CPL, in this case the new constraints for the remaining cosmological parameters are still compatible with their $w$ constant model values.
To generalize more but also check the effect of introducing more degrees of freedom, we allow now the presence of two dark energy components, fluid alike $\Omega_{\rm DE}$ and constant alike $\Omega_{\rm CC}$ as introduced in Sect.~\ref{sec:datamod} and consider the same variants as previous for the EoS, i.e. with the CPL, LZ or $w$ constant parameterisations, while showing minimal extension, or only varying $w_a$, for each. We see in Fig.~\ref{fig:minimalw0wanogeneralisedWCCnoMprio} for all the cases that $\Omega_{\rm CC}$  is close to $\sim$ 0.7, the value we usually obtain within $\Lambda$CDM. We also observe, though with more skewness for the contours, the same behavior for the other involved parameters, as was the case without an additional $\Omega_{\rm CC}$. As so, $w_a$ in the CPL and LZ parameterisation deviate from their null values in compensation to $w_0$ to keep the $w$ constant + $\Omega_{\rm CC}$ (green lines) compatible with $\Lambda$CDM model. And in the minimal scenario where only $w_a$ is varied, the preferred values are the null ones. 

%

\begin{figure}[h]
	\centering
	\includegraphics[width=0.65\columnwidth]{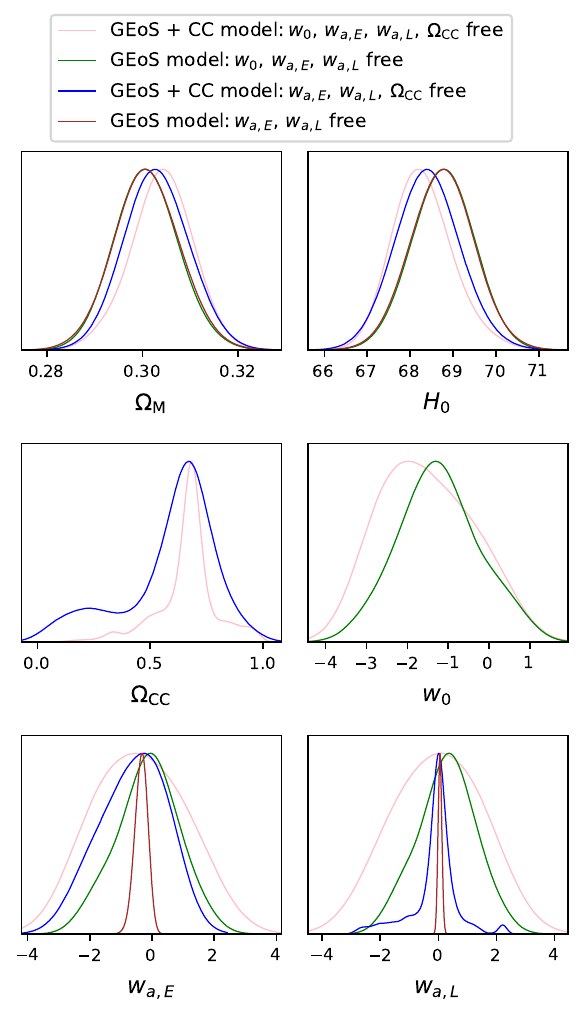}  	
	\caption{The 1D confidence contours marginalised likelihood for the cosmological parameters $\Omega_{\rm M}$ and $H_0$, the fluid-like dark energy equation of state parameters $w_0$, $w_{a,\rm E}$, and $w_{a,\rm L}$ following the GEoS model defined by Eq.~\ref{eq:GEoS}, along with the case where we apply the former model but with further addition of a cosmological constant alike dark energy component as described in Section~\ref{sec:model}, all inferred from a combinations of CMB correlations from Plk18, Supernovae sample from Pantheon+, and BAO measurements from DESI24. }
	\label{fig:minimalw0wanwithgeneralisedWCCnoMprio}
\end{figure}

We end by showing in Fig.~\ref{fig:minimalw0wanwithgeneralisedWCCnoMprio} an ultimate comparison between the GEoS confidence contours  in the context of two fluids, in comparison with the same general parameterisation but with a single dark energy component, along with the same scenario but by fixing $w_0$ to -1 in the two former cases.  We observe, as we expected, that fixing $w_0$ is enough to break degeneracies even in the cases with the most degree of freedom since $w_{a,\rm E}$ and $w_{a,\rm L}$ are still strongly constrained (brown lines) around $\Lambda$CDM values in the minimal GEoS parameterisation. Allowing additionally $\Omega_{\rm CC}$ enlarge the bounds on $w_{a,\rm E}$, though still compatible with its null value, while $w_{a,\rm L}$ remains better constrained in the minimal GEoS+$\Omega_{\rm CC}$ case with respect to the GEoS only one but with a free $w_0$.  Therefore, we see that for all scenarios, we found that either $w$ constant or the minimal LZ are the most economical and constrained models compatible with $\Lambda$CDM for the data we considered in this work, and that the presence of $w_0$ results in degeneracies between the EoS parameters and a widening in the confidence contours. 

Finally, to get another quantitative estimation of the preference and comparison between the different models, we calculate the Bayes factor among them. The Bayes factor of two different models ${\cal M}_i$ and ${\cal M}_j$, described by the parameters $\boldsymbol{\theta}_i$ and $\boldsymbol{\theta}_j$ is given by:
\begin{eqnarray}
B_{ij} \equiv \frac{\int d\boldsymbol{\theta}_i\, \pi(\boldsymbol{\theta}_i \vert {\cal M}_i) {\cal L}(\mathbf{x} \vert \boldsymbol{\theta}_i,{\cal M}_i)\,,}{\int d\boldsymbol{\theta}_j\, \pi(\boldsymbol{\theta}_j \vert {\cal M}_j) {\cal L}(\mathbf{x} \vert \boldsymbol{\theta}_j,{\cal M}_j)\,,}\,,
\label{eq:bayesfactor}
\end{eqnarray}
where $\pi(\boldsymbol{\theta}_i \vert {\cal M}_i)$ is the prior for the parameters $\boldsymbol{\theta}_i$ and ${\cal L}(\mathbf{x} \vert \boldsymbol{\theta}_i,{\cal M}_i)$ the likelihood of the data given the model. A Bayes factor $B_{ij}>1$ (or equivalently $\ln B_{ij}>0$) indicates that model ${\cal M}_i$ is more preferred by data than model ${\cal M}_j$. We adotped the same weak priors used in the DESI24 paper but a different code \cite{Heavens:2017afc} to obtain the different Bayes factor values and we also note that we are using the \texttt{plik-lite} likelihood which is marginalized over the many nuisance parameters of Plk18 \texttt{plik} used by DESI24. Our CMB datasets also does not include Atacama Cosmology Telescope observations \cite{ACT:2023kun} added by DESI24. Nevertheless, the confidence contours we obtain when we consider the same model used in DESI24 still show the same bounds to a high agreement. However, this is only found in our calculations of the Bayes Factor (with our different code ) providing we set the prior volume to unity, e.g. $\vert\ln B_{ij}\vert\sim 0.64$ in favor of $w_0w_a$ over $\Lambda$CDM, while we get different absolute values for $\ln B_{ij}$, albeit with the same expected similar relative overall variation with different scenarios, when we adopt, either the same wide priors as DESI24, or when we consider stronger priors, at the level of 20\% above or below the best fit values. We cautious thus before we continue to the necessity to validate or standardize in the future the few codes that allow to calculate the Bayes Factor from MCMC outputs and we emphasize on the carefulness needed when conducting and comparing these calculations. 

Returning to our results, we found when adopting the same wide priors as DESI24 $\ln B_{ij} \sim 2.65$ when confronting CPL model to $\Lambda$CDM,  indicating a moderate preference for the latter on Jeffreys’ scale \cite{John:2002gg}, not far as expected from the value we get when confronting the LZ model to $\Lambda$CDM. As mentioned above, this preference, is reduced when adopting strong priors, as expected and argued in \cite{2024PhRvD.110l3522A},  to $\ln B_{ij} \sim 0.9$, before it becomes in favor of CPL when setting the prior volume to one. Nevertheless, in all settings, we found that CPL is favored over the more minimal models, with, in the weak priors settings (or even relatively the same trend when using strong priors), $\ln B_{ij} \sim 3.36$  for the $\Lambda$CDM model over $w$ constant, which translates into a moderate preference of CPL over $w$ constant. While for the minimal models, we find for the minimal LZ $\ln B_{ij} \sim 3.92$ and $\ln B_{ij} \sim 2.67$ for the minimal CPL model showing moderate to strong preference for $\Lambda$CDM. However, when confronting to GEoS the generalized parameterisation model, despite the extra degrees of freedom now and the increase in the space of variation, we found strong preference over $\Lambda$CDM and a fortiori over the previous models, with $\ln B_{ij}$ shooting to $> 25$. The effect is also present albeit much smaller when introducing further extensions such as $\Omega_{\rm CC}$ as additional component, where we found that it weakens the preferences for $\Lambda$CDM. with respect to what was the case for either $w$, CPL or the minimal models. As so, we found now only $\ln B_{ij} \sim 0.36$ for $\Lambda$CDM over CPL+CC, close but different now from $\ln B_{ij} \sim 0.51$ for LZ+CC case since the presence of $\Omega_{\rm CC}$ does not allow the recast CPL into LZ, $\ln B_{ij} \sim 0.63$ for $w$+CC,  and $\ln B_{ij} \sim 0.51$ for minimal LZ+CC and  $\ln B_{ij} \sim 0.35$ for minimal CPL+CC, such that we see an almost relative equal variation for all cases. As mentioned, we checked that the above trend stays relatively the same when we set much stronger priors or if we set the prior volume to one such that we are rather evaluating instead a likelihood ratio \citep{Kerscher:2019pzk}. To summarize, we say that the evidence for CPL or LZ are stronger than those for the more minimal cases, however the preference increases further for the more generalized parameterisations instead of penalizing the addition of the extra degrees of freedom, while still staying compatible with $\Lambda$CDM in terms of the significance levels. This corroborates that $\Lambda$CDM is still favored by the datasets we used over the minimal extensions we considered here, and calls for further investigations when testing CPL and models with higher degrees of freedom to converge between preferences from confidence contours and other quantitative statistical tests. 
\section{Conclusion}\label{sect:conclusions}
In this work we wanted to test whether the evidence for dynamical dark energy recently found when combining CMB angular power spectrum, Supernovae luminosity distance measurements, and BAO signature in galaxy clustering catalogs from the DESI collaboration is biased by the type of parameterisation adopted for the dark energy equation of state evolution, namely CPL. Since the latter has the feature of becoming more effective at early times, we proposed to investigate and compare with another parameterisation for the EoS but that has supposedly an impact at late time, namely LZ. For that we perform a bayesian analysis of the two parametrisations using CMB from Planck 2018 and luminosity distance from Supernovae Pantheon+ sample and baryonic acoustic oscillations signature on galaxy clustering from the latest DESI survey release. We also performed similar runs adopting a constant EoS parameter $w$ as a benchmark since it was found that its preferred value is still -1, or what is equivalent to a cosmological constant in LCDM. We found that the same as is the case with CPL model, evidence for dynamical dark energy within the LZ model, but for values almost symmetrically distributed with respect to the $\Lambda$CDM limits of these values i.e. $w_0=-1$ and $w_a=0$. This resulted from the fact that, the presence of $w_0$ allows to recast one model into another, and that each model is basically compensating the preference for the effective $w$ to -1 in the opposite direction in the parameter space with respect to the other model. 
To further test our hypothesis, we had to fix the value of $w_0$ to its $\Lambda$CDM limit and investigate variants where we follow a minimal approach by considering models that deviates only by one free parameter within the same previous parameterisations, i.e what we can call a minimal LZ or a minimal CPL model, with the two models becoming in this case totally different from each others.  We found then that $w_a$ shows no statistically significant shifting and are fully compatible with their null values. To test further the robustness of our findings, we opened or increased the degrees of freedom by extending to more general parameterisations, such as those incorporating both early and late type effect we name GEoS model, or other class of models that allow the presence of two dark energy components, fluid like $\Omega_{\rm DE}$ and constant alike $\Omega_{\rm CC}$. We found, the same as was the case without the latter further extensions, a compatibility for the GEoS parameters with their $\Lambda$CDM limit values, or, for the second class of models, an $\Omega_{\rm CC}$ likelihood peaking at $\sim 0.68$ , the value we obtain for $\Omega_{\rm DE}$  when considering only $\Lambda$CDM. 
When comparing the Bayes factor, we found that the evidence from CPL or LZ are stronger than those for the more minimal cases. However the preference increases further for the more generalized parameterisations, instead of penalizing the addition of the extra degrees of freedom, while still staying compatible with $\Lambda$CDM in terms of the significance levels. We conclude that considering CPL model is not sufficient on its own to test deviations from the standard model and that it is necessary to conduct further minimal or more general approaches, i.e. lowering or increasing the degree of freedom, when confronting models to observations, so that to localize biases, if they exist, or to better understand the outcomes from model testing and inference methods.
 
 \begin{acknowledgments}
Z.S. would like to thank Savvas Nesseris for fruitful discussions and crucial comments on the present work. Z.S. acknowledges support from the IRAP Toulouse and IN2P3 Lyon computing centers.
\end{acknowledgments}

\bibliography{apssamp}

\end{document}